\documentclass[preprints,article,accept,moreauthors]{Definitions/mdpi}

\usepackage{textcomp}
\usepackage{wasysym}
\usepackage{float}
\usepackage{amsmath,amssymb,amsthm}
\usepackage{mathtools}
\usepackage{mathdots}
\usepackage{yhmath}
\usepackage{cancel}
\usepackage{color}
\usepackage{siunitx}
\usepackage{array}
\usepackage{gensymb}

\usepackage{tabularx}
\usepackage{multirow}
\usepackage{booktabs}
\usepackage{pdfpages}

\usepackage{pgfplots}
\pgfplotsset{compat=1.14}
\usepackage{tikz}
\tikzset{every picture/.style={line width=0.75pt}} 

\usepackage[noabbrev, capitalize]{cleveref}

\theoremstyle{remark}
\newtheorem{observation}{Observation}

\firstpage{1} 
\pubvolume{xx}
\issuenum{1}
\articlenumber{5}
\pubyear{2020}
\copyrightyear{2020}
\history{}

\Title{Performance Analysis of Identification Codes}

\Author{Sencer Derebeyo\u{g}lu $^{1}$, Christian Deppe $^{1}$\orcidB{} and Roberto Ferrara $^{1,\ast}$\orcidC{}}

\AuthorNames{Sencer Derebeyo\u{g}lu, Christian Deppe and Roberto Ferrara}

\address[1]{%
	$^{1}$ \quad Lehr- und Forschungseinheit f\"ur Nachrichtentechnik,
	Technische Universit\"at M\"unchen, Munich, Germany; christian.deppe@tum.de, roberto.ferrara@tum.de
}

\abstract{In this paper we analyse the construction of identification codes. Identification codes are based on the question "Is the message I have just received the one I am interested in?", as opposed to Shannon's transmission, where the receiver is interested in not only one, but any message. The advantage of identification is that it allows rates growing double exponentially in the blocklength at the cost of not being able to decode every message, which might be beneficial in certain applications. We focus on a special identification code construction based on two concatenated Reed-Solomon codes and have a closer look at its implementation, analysing the trade-offs of identification with respect to transmission and the trade-offs introduced by the computational cost of identification codes.}

\keyword{Identification, Coding, Construction, Reed-Solomon, Simulation}

\begin{document}
	
	
	\section{Introduction}
	Shannon's transmission~\cite{Shannon} is the standard model for communication. For some special tasks however, more effective models can be used. After Shannon's transmission theory, there have been some proposals on achieving a better capacity, using schemes that are outside of Shannon's description. These advanced techniques for communications are sometimes referred to as Post-Shannon communication models. In 1989, Ahlswede and Dueck proposed one such new model for communication: identification~\cite{AD89,AD89feedback}. The main differences between transmission and identification are the following questions: The receiver in transmission is interested in "What is the message I just received?", whereas the receiver in identification is merely interested in "Is the message I have just received the one I am interested in?" (\cref{fig:question-transmission,fig:question-identification}). In Shannon's transmission, there is usually one receiver and he is ready to decode any codewords in the codebook he receives into messages. However, there can be multiple receivers in identification and each of them would only be interested in one message of the codebook.
	
	\begin{figure}
		\centering
		\tikzset{every picture/.style={line width=0.75pt}} 
		
        \begin{tikzpicture}[xscale=2.5,
		    box/.style={draw, thick, align=center, minimum height=1.25cm, minimum width=2cm}]
		\path (0,0)
    	    node[box] (source) {Sender}
		    ++ (2,0)
		    node[box] (channel) {\quad Noisy Channel\quad }
		    ++ (2,0)
		    node[box] (sink) {Receiver}
		    ;
        
        \draw[->] (source)  -- (channel);
        \draw[->] (channel) -- (sink);
        
        \node[below of=source] {sends message $i$};
        \node[below of=sink] {What is the message?};
        \node[above of=sink] {interested in any message};
        
        \end{tikzpicture}		

		\caption{An illustration of a transmission receiver interested in any message.}\label{fig:question-transmission}
		
		\bigskip 

        \begin{tikzpicture}[xscale=2.5,
		    box/.style={draw, thick, align=center, minimum height=1.25cm, minimum width=2cm}]
		\path (0,0)
    	    node[box] (source) {Sender}
		    ++ (2,0)
		    node[box] (channel) {\quad Noisy Channel\quad }
		    ++ (2,0)
		    node[box] (sink) {Receiver}
		    ;
        
        \draw[->] (source)  -- (channel);
        \draw[->] (channel) -- (sink);
        
        \node[below of=source] {sends message $i$};
        \node[below of=sink] {Did he send me message $j$?};
        \node[above of=sink] {interested only in message $j$};
        
        \end{tikzpicture}

		\caption{An illustration of an identification receiver only interested in one message.}\label{fig:question-identification}
	\end{figure}
	
	A real life application of identification can be online sales. Since web platforms track user data in their surfing or product viewing behaviours, they can identify whether the user is interested in a certain product type or company. Various categories are possible for data tracking. We can think of each element in these categories as a receiver of an identification scheme. According to the information of the user's interests gathered with the help of identification, the platform can use optimised advertising when targeting that user~\cite{BD18wiretapID}.
	
	There are advantages and disadvantages to identification. It allows a negligible amount of overlapping between the decoding sets of codewords. This helps to fit in exponentially more messages~\cite{AD89,AD89feedback}, but the drawback is that we cannot decode every message anymore. The additional error introduced by the overlap can still be made arbitrarily close to zero in the asymptotic case, namely for codewords which have increasingly large blocklengths.
	
	Identification codes can be constructed around transmission codes with the addition of some pre-processing at the sender and some post-processing at the receiver.
	Rather than associating a codeword and a decoding set to each identification message, or \emph{identity}, the identification scheme associates a function to each identification message.
	Given an identity, the identification encoder then picks an input to the function at random, computes the corresponding output and sends the input-output pair to the transmission encoder so that it can be sent through the channel. 
	Given any other identity, the identification decoder first gets an estimate for the input-output pair from the transmission decoder, and then it computes its own output on the received input using the function associated to the given identity.
    The identity given to the identification decoder is verified if the computed output matches the received output and rejected otherwise. 
    Beyond the errors that can be introduced by the transmission code use to send the input-output pair, an error can occur also when transmission was successful if the two identities provided to the identification encoder and decoder are distinct, but nonetheless produce the same output on the randomly chosen input.
    In such case the identity at the decoder would be erroneously accepted as verified.
    
    The identities given to the identification encoder and decoders are uniquely determined by their associated function. 
    One of the goals of constructing identification codes is to find functions distinct enough so that the probability of erroneously veryfing an identity in the manner just described is negligible.
    These functions turn out to be simply error-correction codes that are used in a different manner than how they are used for the correction or error in transmission schemes.
    In order to achieve identification capacity some conditions must be satisfied~\cite{VW93explicit}, and the existance of codes satisfying this condition is guaranteed by either probabilistic arguments~\cite{AD89feedback} or by the Gilbert-Varshamov bound~\cite[Section~III.B]{AZ95newdirections}\cite[Appendix]{VW93explicit}.
	In this paper, we focus on a capacity-achieving construction of identification codes based on the concatenation of two Reed-Solomon codes\cite{VW93explicit}. 
	The codewords made with the help of the concatenated Reed-Solomon codes are those mapping functions we have mentioned above. The positions of the symbols in the codewords are the inputs and the outputs are nothing but the symbols themselves.

    The paper is structured as follows. 
    In the next section (\cref{chap:preliminaries}) we introduce all the necessary concepts for an implementation of the identification code based on the concatenated Reed-Solomon code.
    For further details on identification we refer also to an upcoming survey of identification~\cite{survey}.
	In \cref{chap:simulation} we describe how we compare such construction to transmission and how the simulation is implemented, with particular focus on the efficiency of the implementation.
	We also give an analytic formula for how much larger is the growth of number of identities compare to the number of transmission messages.
    Finally, in \cref{chap:results} we discuss the numerical results obtained in the simulation.
	
	\section{Preliminaries}
	\label{chap:preliminaries}
	
	Channels are mediums in which the information-carrying codewords are transmitted. Noisy channels disrupt the codewords they are carrying, unless we are using an ideal channel. 
	

	\begin{figure}
		\centering      

        \begin{tikzpicture}[xscale=1.7,
		    box/.style={draw, thick, align=center, minimum height=1.25cm, minimum width=1.5cm}]
		\path (0,0)
    	    node[box] (source) {Source}
		    ++ (2,0)
		    node[box] (enc) {Transm.\\Encoder}
		    ++ (2,0)
		    node[box] (channel) {Noisy\\Channel}
		    ++ (2,0)
		    node[box] (dec) {Transm.\\Decoder}
		    ++ (2,0)
		    node[box] (sink) {Sink}
		    ;
        
        \draw[->] (source)  -- node[above,font=\footnotesize] {message}    (enc);
        \draw[->] (enc)     -- node[above,font=\footnotesize] {codeword}   (channel);
        \draw[->] (channel) -- node[above,font=\footnotesize] {codeword$'$}(dec);
        \draw[->] (dec)     -- node[above,font=\footnotesize] {message$'$} (sink);
        
        \end{tikzpicture}		
		
		\caption{General transmission scheme} \label{fig:classicaltransmission}

		\bigskip

        \begin{tikzpicture}[xscale=1.7,
		    box/.style={draw, thick, align=center, minimum height=1.25cm, minimum width=1.5cm}]
		\path (0,0)
    	    node[box] (source) {Source}
		    ++ (2,0)
		    node[box] (enc) {ID\\Encoder}
		    ++ (2,0)
		    node[box] (channel) {Noisy\\Channel}
		    ++ (2,0)
		    node[box] (dec) {ID\\Decoder}
		    ++ (2,0)
		    node[box] (sink) {Sink}
		    ;
        
        \draw[->] (source)  -- node[above,font=\footnotesize] {identity}    (enc);
        \draw[->] (enc)     -- node[above,font=\footnotesize] {codeword}   (channel);
        \draw[->] (channel) -- node[above,font=\footnotesize] {codeword$'$}(dec);
        \draw[->] (dec)     -- node[above,font=\footnotesize] {yes/no} node[below,font=\footnotesize] {accept/reject} (sink);

        \draw[<-] (enc) -- node[left,font=\footnotesize, align=center] {Local\\Randomness} ++ (0,-2);
        \draw[<-] (dec) -- node[left,font=\footnotesize, align=center] {Identity$'$} ++ (0,-2);

        \end{tikzpicture}

		\caption{General identification scheme where local randomness at the encoder is necessary to achieve capacity, as opposed to the transmission encoder where it is optional.} \label{fig:identicaltransmission}
		
	\end{figure}

	Shannon's transmission, is a system where a party communicates messages to another via an often noisy channel. The receiver is interested in any message the sender sends, therefore the receiver's motivation can be efficiently stated with the question: "What is the message?"

	To counteract the noise of the channel, the messages are channel-coded into codewords, adding redundancy. Each message is encoded by an encoder before the channel and, after it passes through the channel, the codeword is decoded by a decoder (\cref{fig:classicaltransmission}). The decoder is defined by decoding sets, one for each possible message, which are disjoint\footnote{Or at least the size of the intersection is bounded as the blocklength $n$ increases in the case of soft decoding.}. If a codeword falls into a decoding set after passing through the channel, the decoder recognises the message associated to the decoding set and outputs that particular message. In this way the message has been decoded and transmitted to the sink, which is the communication partner waiting on the receiving end.

	\subsection{Identification}
	Identification differs from Shannon's transmission in how it carries information between two communication partners. 
	This time the receiver is not interested in finding which message $i$ was sent, but only whether one of them, e.g. in message $j$ chosen by the receiver itself, was sent or not.
	The identification messages can be called identities in order to distinguish them from the transmission messages.
	Compared to transmission, the role of the decoder then becomes that of a verifier which either accepts or rejects the received output as being compatible with the chosen message (\cref{fig:identicaltransmission}).

	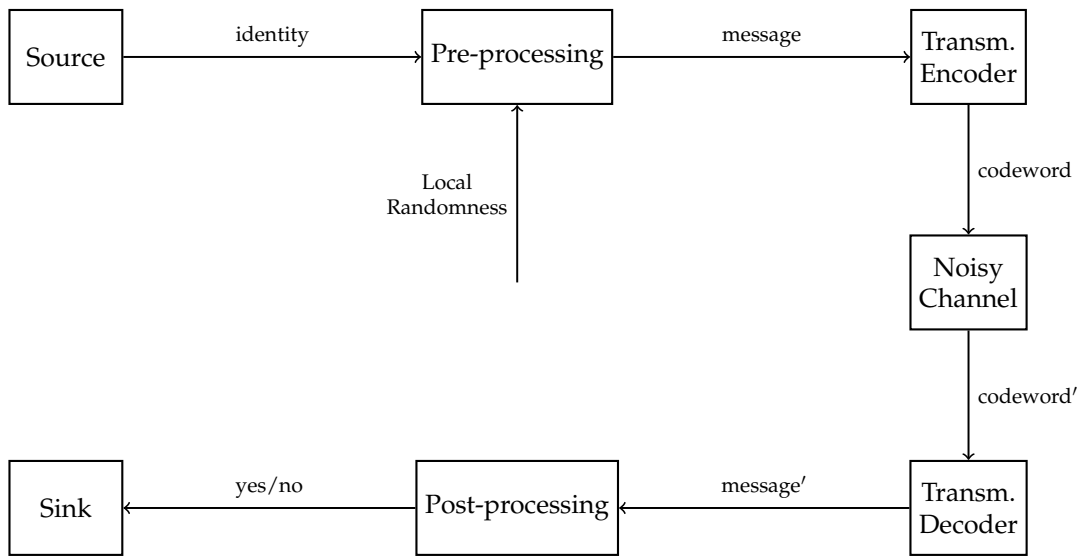
\begin{figure}
		\centering

        \begin{tikzpicture}[xscale=3, yscale=1.5,
            box/.style={draw, thick, align=center, minimum height=1.25cm, minimum width=1.5cm}]
		\path (0,0)
    	    node[box] (source) {Source}
		    ++ (2,0)
		    node[box] (preenc) {Pre-processing}
		    ++ (2,0)
		    node[box] (enc) {Transm.\\Encoder}
		    ++ (0,-2)
		    node[box] (channel) {Noisy\\Channel}
		    ++ (0,-2)
		    node[box] (dec) {Transm.\\Decoder}
		    ++ (-2,0)
		    node[box] (postdec) {Post-processing}
		    ++ (-2,0)
		    node[box] (sink) {Sink}
		    ;
        
        \draw[->] (source)  -- node[above,font=\footnotesize] {identity}    (preenc);
        \draw[->] (preenc)  -- node[above,font=\footnotesize] {message}     (enc);
        \draw[->] (enc)     -- node[right,font=\footnotesize] {codeword}    (channel);
        \draw[->] (channel) -- node[right,font=\footnotesize] {codeword$'$} (dec);
        \draw[->] (dec)     -- node[above,font=\footnotesize] {message$'$}  (postdec);
        \draw[->] (postdec) -- node[above,font=\footnotesize] {yes/no}      (sink);

        \draw[<-] (preenc) -- node[left,font=\footnotesize, align=center] {Local\\Randomness} ++ (0,-2);

        \end{tikzpicture}

		\caption{The specific class of identification schemes which is the focus of this paper.}\label{fig:IdoScheme}
	\end{figure}
	
	Identification can be achieved by adding some pre-processing and post-processing around transmission (\cref{fig:IdoScheme}). These new steps allow for a much larger number of possible identities than the possible transmission messages, while adding a small new kind of error. The idea is that the high number of identities can be beneficial if we can keep the new kind of error below a certain threshold, and send it to zero in the asymptotic case.
	This also allows us to first fix the errors in the channel using a transmission code, thus approximating a noiseless channel, and then code for identification.
	For this reason, it is enough to study identification for the noiseless channel, which is the case for this work.

	
	The advantage of identification over transmission, given the added complexity, lies in the number of possible identities in an identification scheme. By allowing the decoding sets to collide, the maximum number of identities grows doubly exponentially, while the number of messages in the transmission scheme grows only exponentially.
	In short, in identification, we allow a small error to happen and we sacrifice decoding for an extremely large number of identities.
	
	\subsubsection{Types of Errors}	
	In identification we have to bound two kinds of errors independently: missed identification, also known as the error of the first kind, and false identification, also known as the error of the second kind.
	
	The missed identification is a regular transmission error, i.e. the codeword $i$ is not recognised as codeword $i$ on the receiver end. 
	This happens when the receiver is interested in the same message $i$ sent by the sender, and when the message is disrupted in the channel in such a way that the receiver doesn't identify the received output as message $i$. This error is absent when coding for the noiseless channel with the identification codes that we consider.
	
	The false identification will be our main interest in this paper. If the sender sends the message $i$ and the receiver is only interested in a \emph{different} message $j$, the receiver may still identify the message as message $j$. This is a false identification or a false identification error. There are two possible contributions to this error probability:
	
	\begin{itemize}
		\item A transmission error (the same reason as for the missed identification) happens and the channel gives an output of message $j$ although having received the message $i$,
		
		\item The scheme uses a randomised identification code (see \cref{Randomized}) and the randomly selected codeword lies in the overlapping section of the decoding sets of message $i$ and message $j$.
	\end{itemize}
	Since we only look at identification over the noiseless channel, the first contribution is absent.

	\subsubsection{Identification Codes} \label{det} \label{Randomized}
	
	A deterministic code is a sequence of encoding codewords and decoding sets, or better \emph{verifiers}, $\{u_i,D_i\}_{i=1,\dots,N}$ for each possible identity. 
	A randomised code is a sequence of encoding probability distributions and verifiers $\{Q_i,D_i\}_{i=1,\dots,N}$ for each possible message. 
	The randomisation is necessary to achieve a high number of identities in the discrete memoryless channel~\cite{AD89}, which is in contrast to transmission, where having local randomness doesn't increase capacity~\cite{AW69}.
	For this reason we focus only on the randomised codes.
	The construction we work on in this paper also makes use of a randomised identification code. 
	
	In both cases the verifier sets are subsets of the possible outputs, namely {a verifier set $D_i$ is a subset of the $Y^n$ alphabet, where an output codeword is recognised as the $i$-th codeword, if it is an element of $D_i$. The larger the decoding sets are, the more tolerance the code has against transmission errors.}
	The difference between the verifier sets of identification and the decoding sets of transmission is that the do not need to be disjoint, and a priori no restriction is put on the intersection between two distinct verifier sets.
	
	As already mentioned before, we have to treat two types of error independently.
	Indeed, in the constructions sketched in the introduction, we have seen that an error in the post-processing can happen even if there was no error in the transmission. 
	$(n, N, \lambda_1, \lambda_2)$ randomised identification codes are defined as randomised codes with blocklength $n$, $N$ total number of identities and $\lambda_1$, $\lambda_2$ bounds on the two types of errors. Namely, the codewords on the sender side are selected randomly, using a conditional probability distribution $Q$ such that:
	\begin{align}
		&\mu_1^{i} = \sum_{x_n \in X^n}Q(x_n|i)W^n(D_i^c|x^n) \leq \lambda_1, \ \forall i\\
		&\mu_2^{i,j}= \sum_{x_n \in X^n}Q(x_n|j)W^n(D_i|x^n) \leq \lambda_2, \ i \neq j
	\end{align}
	where $Q_i \in X^n$ represents the encoder of the $i$-th identity and $D_j  \subset Y^n$ is the verifier set on the receiver side for the $j$-th identity.
	$\mu_1$ are the probabilities of the error of the first kind and $\mu_2$ are the probabilities of the error of the second kind.
	{Note that the error of the first kind concerns only one codeword, where the error of the second kind happens between two codewords.}
	$\lambda_1$ is thus a bound on the maximum probability of the error of the first kind and $\lambda_2$ is a bound on the maximum probability of the error of the second kind.
	
	As a final note, we point out that, at least for false identification, the error to be bound must be the maximum and not the average error.
	It has been proven that bad codes exist if we require only the average of the false identification probability to be small, these codes allow for an infinite identification rate~\cite{HV92newresults}.

	\subsubsection{Identification codes from Transmission codes} \label{section:tag-codes}
	
	A common way of constructing identification codes is the way explained in the introduction, by having all the channel errors corrected by a transmission code and performing identification for the almost-noiseless channel provided by the transmission code.
	In the noiseless channel then, identification can be done by associating to each identity $i$ a function $T_i$ and by sending a single input-output pair $(j,T_i(j))$ for verification, with $j$ picked uniformly at random using the local randomness highlighted in \cref{fig:identicaltransmission,fig:IdoScheme}, through the channel. 
	The receiver, given another identity $i'$, can verify whether the two chosen identities are the same by computing its own output $T_{i'}(j')$ on the input part $j'$ of the received input-output pair $(j',t')$, and verifying that the received and computed output are the same ($T_{i'}(j')=t'$?). 
	If the channel is noiseless, then $(j',t')=(j,T_i(j))$ this scheme has zero missed identification error (if $i=i'$ then the computed outputs will match) and false identification bounded by the fraction of inputs any two identities map to the same output (if $i\neq i'$ then an incorrect accept will happen only on those $j$ such that $T_{i'}(j)=T_{i}(j)$)~\cite{AD89feedback,VW93explicit}.
	In case of a noisy channel, the error probability of the transmission code (the probability that $(j',t')\neq(j,T_i(j))$) adds to the missed- and false-identification error probabilities of the identification code~\cite{AD89,VW93explicit}.
	Arguably, in such cases the transmission code should have the error probability comparable to the error probability of the identification code. This analysis is however left for future work.

	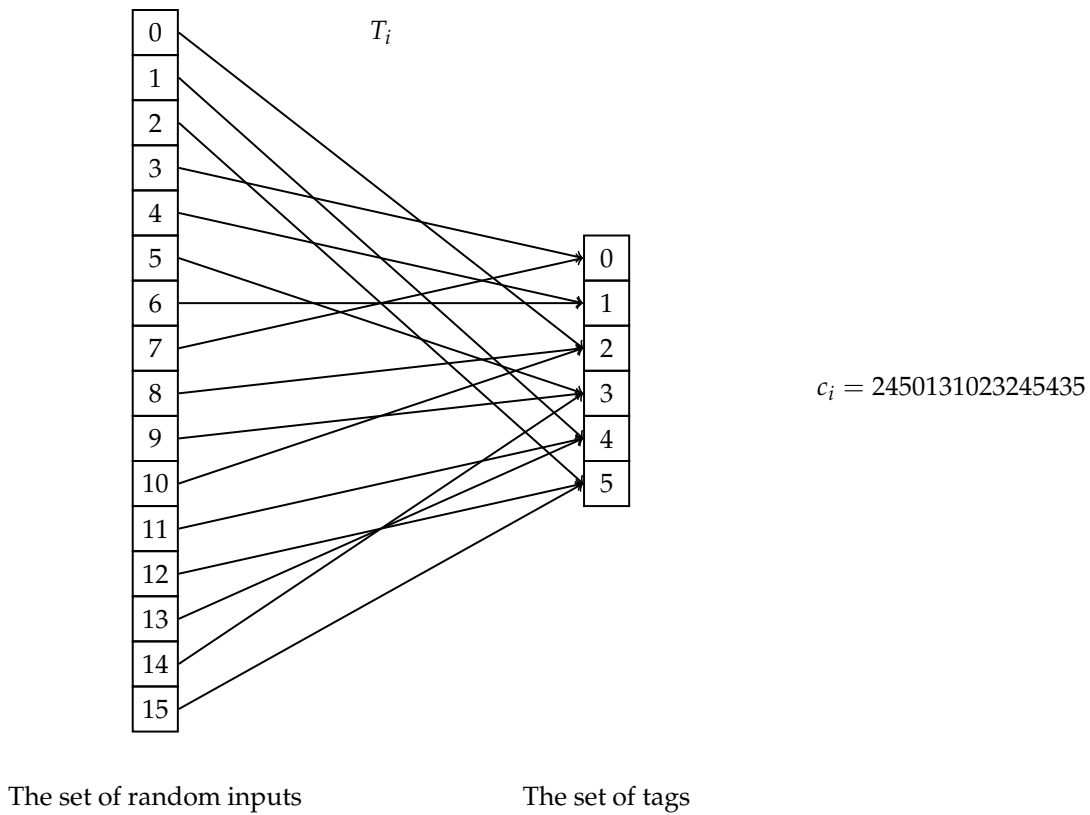
\begin{figure}
		\centering
		\tikzset{every picture/.style={line width=0.75pt}} 
		
		\begin{tikzpicture}[scale=.6]
		\path (0,5) 
		    foreach \x in {0,1,...,15} {
    		    node[draw, minimum width=.6cm, minimum height=.6cm] (r\x) {\x}
    		    ++ (0,-1)
		    }
		    ++ (0,-1) node {The set of random inputs}
		    ;

		\tikzset{shift={(10,0)}}
		
		\path (0,0) 
		    foreach \x in {0,1,...,5} {
    		    node[draw, minimum width=.6cm, minimum height=.6cm] (t\x) {\x}
    		    ++ (0,-1)
		    }
		    ++ (0,-6) node {The set of tags}
            ;
		
		\node at (-5,5) {$T_i$};

		\draw[->] (r0.east)  -- (t2.west);
		\draw[->] (r1.east)  -- (t4.west);
		\draw[->] (r2.east)  -- (t5.west);
		\draw[->] (r3.east)  -- (t0.west);
		\draw[->] (r4.east)  -- (t1.west);
		\draw[->] (r5.east)  -- (t3.west);
		\draw[->] (r6.east)  -- (t1.west);
		\draw[->] (r7.east)  -- (t0.west);
		\draw[->] (r8.east)  -- (t2.west);
		\draw[->] (r9.east)  -- (t3.west);
		\draw[->] (r10.east) -- (t2.west);
		\draw[->] (r11.east) -- (t4.west);
		\draw[->] (r12.east) -- (t5.west);
		\draw[->] (r13.east) -- (t4.west);
		\draw[->] (r14.east) -- (t3.west);
		\draw[->] (r15.east) -- (t5.west);
		
		\node at (8,-2.5,) {$c_i = 2450131023245435 $};
		\end{tikzpicture}

		\caption{An example of a mapping function $T_i$ which belongs to the identity $i$. On the right, the 16-digit-long codeword $c_i$ obtained by concatenating all the function outputs.}\label{fig:mappingfunction}
	\end{figure}
	
    \begin{observation}
    A set of $N$ functions $\{T_i\}_{i=1,\dots,N}$ from inputs of size $M$ to outputs of size $q$ such that, for any two of these functions their outputs are equal on at most $M-d$ inputs, corresponds to an error correction code of size $N$, blocklength $M$, alphabet of size $q$ and distance $d$, and vice versa.
    To each function corresponds a codeword, obtained by concatenating all the possible function outputs in order as displayed in the example of \cref{fig:mappingfunction}.
    Similarly, given any correction code, each codeword defines a function, namely the function that map positions (the size of the blocklength) to the symbols of the codeword.
    \end{observation}
    
    For the sake of clarity, we repeat the identification pre- and post-processing using error-correction codes.
    Given an error-correction code of size $N$, blocklength $n$, alphabet of size $q$ and minimum Hamming distance $d$, namely a $(n,N,d)_q$ error-correction code, an identification code is constructed as follows.
    The error-correction code is not used to correct error, but instead is used in a different manner.
	The codewords $c_i$ of the error-correction code are associated each to an identity $i$. 
	The identification sender of identity $i$ randomly and uniformly chooses a position $j$ from $1,2,\dots,n$, and then sends $j$ and the $j$th letter $c_{ij}$ to the receiver, using a transmission code if the channel is noisy.
	The receiver must make a choice on what identity he is interested, say $i'$.
	Upon receiving $j,c_{ij}$, the receiver checks whether $j$th letter of the codeword associated with his interested message $i'$th is $c_i$.
	Namely, it checks whether $c_{i'j} = c_{ij}$ and says ``yes/accept $i'$'' if so, or ``no/reject $i'$'' otherwise. 
	In the noiseless case, the only possible error is the false identification error (error of the second kind), which occurs only when the receiver is interested in a different message than the one sent, and two codewords $c_i$ and $c_{i'}$ have the same letter as in the $i$th position. Thus, the probability of false identification is bounded by 
	\begin{equation}
	    \label{eq:ecc-falseIDerror}
	    \lambda_2 \leq 1-\frac{d}{n}.
	\end{equation}
	Again, we highlight that this is a bound on the false identification error only in the absence of a transmission error.

	In order to uniquely refer to the input-output pairs $(j,T_i(j))$ produced in the pre-processing we will call $j$ the \emph{randomness} and $T_i(j)$ the \emph{tag}, for reason that should become clear next.
	For convenience we may then call the functions $T_i$ \emph{tagging functions}.
	Traditionally, the identities have been called messages~\cite{AD89,AD89feedback}.
	However, the identities are not the messages that are sent through the transmission code.
	Furthermore, transmission can be performed in parallel to identification without trade-off, meaning that the capacity of both can be achieved at the same time~\cite{HV92newresults}.
	This is because the goal of the pre-processing is to use all the capacity of the channel to send the local randomness $j$ and produce common randomness between the sender and receiver~\cite{AC06}.
	A small tag size (the size of $T_i(j)$ or $c_{ij}$) that asymptotically does not use any capacity~\cite{AD89feedback} is then enough to allow a rate of identities that grows doubly exponentially in the blocklength.
	The intuition is that identification is performed by verifying the two tags, the senders and the receivers, of a random challenge.
	In other works~\cite{BD18wiretapID,BD19jammingID,BDW19}, the randomness has been called a colouring number, the tag a colour and the identity a colouring.

	\begin{observation}
    In~\cite{VW93explicit,KY99hashID} error-correction codes are used to construct constant-weight binary codes that are then used to construct identification codes following the method in~\cite{VW93explicit}.
    This at first glance might seem like a different way of constructing identification codes, however it is implicitly the same use of error-correction codes presented above.
    The way the constant-weight binary codes are further encoded in~\cite{VW93explicit} (via the use of the incidence matrix of the binary code), its an implicit encoding of the randomness-tag pair.
    Both in~\cite{VW93explicit,KY99hashID} and in the scheme above, the information that is sent through the channel via the transmission code is a randomness-tag pair.
	\end{observation}

	\subsection{Error-Correction Codes} \label{ECC}
	
	We have seen error-correction codes used for identification.
	However, their initially intended purpose is to allow the receiver to fix the damage a noisy channel does to the codewords, by adding some redundancy to the original message.
	This intended purpose has a huge influence on how the encoding is generally performed. 
	The result is that computing the codewords in the usual way (using the generator matrix) and then computing a tag for identification as explained above results in an extremely inefficient implementation. 
	For this reason we briefly review the background in error-correction codes and Reed-Solomon codes in particular, in  order to highlight were we obtain the more efficient implementations.
	
	Linear block codes are error-correction codes over an alphabet such that any linear combination of its codewords is also a codeword.
	The $[n,k,d]_q$ codes denote the set of linear codes over an alphabet of size $q$ (commonly omitted for $q=2$), mapping length $k$ message to blocklength $n$ codewords of minimum Hamming distance $d$. The size of the a block code is $q^k$.
	The linearity of the code implies that we can see messages as $k$ dimensional vectors, that  we can build a generator matrix $G$ and compute any codeword for any message via matrix multiplication as 
	\begin{equation}
		\underline{m} \cdot G = \underline{c}
	\end{equation}
	where $m$ is the message, i.e. the original bitstream, $G$ is the generator matrix of the linear code and $c$ is the resulting codeword after the encoding process. $G$ will have full rank and the dimensions $k\times n$ in order to map messages to codewords uniquely. 
	
	
		


	\subsubsection{Reed-Solomon Codes} \label{subsect:rs}
	
	Reed-Solomon codes~\cite{RS60} are linear block-codes based on either prime fields or extension fields.
	The key observation of this subsection is that Reed-Solomon codes naturally define tagging functions before constructing the error-correction codewords.
	Reed-Solomon codes also achieve the Singleton Bound 
	\begin{equation}
		d \leq n - k + 1
	\end{equation}
	with equality, which means that they are maximum distance separable codes.
	This is relevant for tag codes, because greater scaling in the code distance means a smaller false identification probability in the asymptotic case.
	
	The codewords of Reed-Solomon codes are generated as the evaluation of polynomials on the elements of $F_q$, also called code locators. To each message corresponds a polynomial, where the symbols of the message are the coefficients of the polynomial. 
	So, if we let $\underline{m} = \{m_0,m_1,m_2,m_3,m_4,...,m_{k-1}\}$ be the message vector and have $\beta \in F_{q}$, then the evaluation of $\beta$ is defined as:
	\begin{equation}
		T_m(\beta) = \sum_{i=0}^{k-1} m_i \cdot \beta^i
	\end{equation}
	with the convention that $0^0=1$ (we use the same codes as \cite{VW93explicit}). For a Reed-Solomon code of length $n\leq q$, there are $n$ such evaluations. 
	One evaluation gives us one symbol of the codeword, whereas $n$ such evaluations give us the whole codeword. If we let $\underline{c} = \{c_0,c_1,c_2,c_3,c_4,...,c_{n-1}\}$ be the codeword and $f_j \in F_q \ \forall i$, where $j$ shows us the order of the picked code locator, then the codewords are:
	\begin{align}
		\underline{c}(m) 
		&= \bigg
		\{\sum_{i=0}^{k-1} m_i \cdot f_0^i, 
		&&\sum_{i=0}^{k-1} m_i \cdot f_1^i,
		&&\sum_{i=0}^{k-1} m_i \cdot f_2^i,
		&& ... ,
		&&\sum_{i=0}^{k-1} m_i \cdot f_{n-1}^i \bigg\}
		\\
		&= 
		\{T_m(f_0) , 
		&&T_m(f_1)  , 
		&&T_m(f_2) , 
		&& ... , 
		&&T_m(f_{n-1})\}.
	\end{align}
	These codewords can be generated from the message $m=m_1 \dots m_{k-1}$ with the $k \times n$ generator matrix:
	\begin{align}
		G = \begin{pmatrix}
			f_0^0 & f_1^0 & f_2^0 & f_3^0 & f_{n-1}^0\\
			f_0^1 & f_1^1 & f_2^1 & f_3^1 & f_{n-1}^1\\ 
			f_0^2 & f_1^2 & f_2^2 & f_3^2 & f_{n-1}^2\\
			f_0^3 & f_1^3 & f_2^3 & f_3^3 & f_{n-1}^3\\
			\\
			\vdots  &  & \ddots & & \vdots\\
			\\
			f_0^{k-1} & f_1^{k-1} & f_2^{k-1} & f_3^{k-1} & f_{n-1}^{k-1}
		\end{pmatrix}.
	\end{align}
	Two distinct polynomials of degree less than $k$ agree on at most $k-$ inputs, therefore the distance of the code constructed this way in $n-k+1$.
	We denote with
	\[(n,k)_q^{RS}\]
	the $[n,k,n-k+1]_q$ Reed-Solomon block code.
	We will be interested only in the $(q,k)_q^{RS}$ codes
	
	If we use the Reed-Solomon code for identification, then the polynomials $T_m$ are the tagging functions, the randomness is $j\in\{0,\dots,n-1\}$, $F_q$ are the possible tags, and $(j, T_m(j))$ is the randomness-tag pair sent through the transmission code.
	
	

	\subsubsection{Concatenated Codes}
	
	We will see that using a single Reed-Solomon code as tagging functions does not allow to achieve the identification capacity~\cite{VW93explicit}.
	A way around this is to concatenate multiple codes~\cite{VW93explicit}.
	When concatening two codes, we distinguish the two as inner code $\mathcal{C}_\mathrm{i}$ and outer codes $\mathcal{C}_\mathrm{o}$ with the usual convention that the outer code's symbols are encoded by the inner code.
	The subscripts ``$\mathrm{i}$'' and ``$\mathrm{o}$'' stand for the inner and the outer code respectively
	Let the inner code be an $[n_\mathrm{i}, k_\mathrm{i}, d_\mathrm{i}]_{q_\mathrm{i}}$ code and the outer code be an $[n_\mathrm{o}, k_\mathrm{o}, d_\mathrm{o}]_{q_\mathrm{o}}$ code, satisfying $q_\mathrm{i} = q_\mathrm{o}^{k_\mathrm{o}}$. Then the concatenated error-correction code $\mathcal{C}_\mathrm{i} \circ \mathcal{C}_\mathrm{o}$ is an
	\[[n_\mathrm{c}, k_\mathrm{c}, d_\mathrm{c}]_{q_\mathrm{c}} = [n_\mathrm{i}n_\mathrm{o}, k_\mathrm{i}k_\mathrm{o}, d_\mathrm{i}d_\mathrm{o}]_{q_\mathrm{i}}\] code.
	The alphabet of the concatenated code is the alphabet of the inner code $q_\mathrm{i}$. 
	The number of codewords in the concatenated code is the number of codewords in the outer code, $q_\mathrm{i}^{k_\mathrm{i}k_\mathrm{o}} = q_\mathrm{o}^{k_\mathrm{o}}$. 
	
	The following is an example taken from~\cite{krish} of concatenated codes where each symbol of the outer code is encoded by the inner code, albeit without using block codes {(these are however not block codes and thus $k_\mathrm{c} = k_\mathrm{i}\cdot k_\mathrm{o}$ is not valid for this example)}. $\mathcal{C}_\mathrm{i} = \{0120112, 1202102, 2100211, 1201120\}$ with $n_\mathrm{i} = 7$, $q_\mathrm{i} = 3$, $|\mathcal{C}_\mathrm{i}| = 4$ and $\mathcal{C}_\mathrm{o} = \{ad, bc, ac, cc, db, ab\}$ with $n_\mathrm{o} = 2, alphabet A_\mathrm{o} = \{a,b,c,d\}$ ($q_\mathrm{o}=4$), $|\mathcal{C}_\mathrm{o}| = 6$. Then:
	\begin{align}
		\mathcal{C}_\mathrm{i} \circ \mathcal{C}_\mathrm{o} = 
		\{ &0120112\,1201120
		\\ &1202102\,2100211
		\\ &0120112\,2100211
		\\ &2100211\,2100211
		\\ &1201120\,1202102
		\\ &0120112\,1202102\},
	\end{align}
	where $n_\mathcal{c} = 14$, $q_\mathcal{c} = 3$, $|\mathcal{C}_\mathrm{i} \circ \mathcal{C}_\mathrm{o}| = 6$. 
	The outer code's alphabet size and the inner code's number of codewords in the codebook are the same so that each codeword in the inner code represents one symbol in the alphabet of the outer code. E.g.: 0120112 represents $a$, 1202102 represents $b$, 2100211 represents $c$ and 1201120 represents $d$. This is clear in the codebook of the concatenated code.

	\subsection{Error-correction codes achieving identification capacity}
	\label{threeconditions}
	
    There are three conditions which a capacity-achieving identification code constructed on transmission codes needs to meet. 
	If we are considering a sequence of block codes $[M(n),k(n),d(n)]_{q(n)}$ to construct identification codes, the following conditions are necessary in order to achieve the identification capacity~\cite[Definition~9]{VW93explicit}:
	\begin{enumerate}
		\item The size of the block code and thus the size of the identification code must be exponential in the size of the randomness:
		\begin{align}
			\label{eq:optimal-size}
			\lim_{n \to \infty} \frac{\log k(n)}{\log M(n)} &\to 1;
		\end{align}
		\item The size of the tag in bits must be negligible in the size of the randomness in bits:
		\begin{align}
			\label{eq:optimal-tag}
			\lim_{n \to \infty} \frac{\log q(n)}{\log M(n)} &\to 0;
		\end{align}
		\item The distance of the code must grow as the blocklength, so that the error of the second kind goes to zero:
		\begin{align}
			\label{eq:optimal-error}
			\lim_{n \to \infty} \frac{d(n)}{M(n)} &\to 1.
		\end{align}
	\end{enumerate}
	Recall that after computing the tag, the tuple must be channel coded.
	The first two conditions together guarantee that the entire capacity of the channel is used to send the randomness, and that the number of identities grows doubly exponential in the channel capacity. The last condition guarantees that we can send the false identification error to zero.

	The Gilbert-Varshamov bound guarantees the existence of such codes, but it is actually possible to explicitly construct,
	given a capacity achieving code for transmission, identification codes that achieve the identification capacity~\cite{VW93explicit}.
	One such construction involves the concatenation of two Reed-Solomon codes, where one is based on a prime field and the other is based on an extension field. The construction is defined as: 
	\begin{equation}
	    \label{eq:RS2}
	    (q,k,\delta)_{RS^2} \coloneqq (q,k)_q^{RS} \circ (q^k, q^{k-\delta})_{q^k}^{RS},
	\end{equation}
	which gives a
	\begin{equation}
		[q^{k+1}, kq^{k-\delta}, (q-k+1)(q^k-q^{k-\delta}+1)]_q
	\end{equation} 
	block code.
	We will refer to it either a \emph{the double} or \emph{concatenated Reed-Solomon code}.
	
	To achieve the identification capacity as the blocklenght $n$ grows, $q$, $k$ must be increasing functions of $n$ satisfying~\cite[Proposition~3]{VW93explicit} 
	\begin{equation}
	    \label{eq:2RSscaling}
	     q(n) \gg k(n) \gg \delta(n) > 0
	\end{equation}
	($\delta$ can stay constant).
	These conditions allow the concatenated Reed-Solomon code to satisfy the capacity achieving conditions below.
	The first condition, \cref{eq:optimal-size}, can be easily verified. The second condition, \cref{eq:optimal-tag}, is also obvious, because $q^{k+1}$ will always be bigger and also grow faster than $q$ itself. 
	In particular, these two conditions are met already, by expanding the extension field into the base field for the outer code only.
	The crucial condition where the inner code is needed is the third one, \cref{eq:optimal-error}. 
	Applying the code distance of the double Reed-Solomon code construction to the third condition gives:
	\begin{align}
		\frac{d}{M} &=\frac{(q-k+1)(q^k-q^{k-\delta}+1)}{q^{k+1}}
		\geq \frac{(q-k)(q^k-q^{k-\delta})}{q^{k+1}}
		\\&
		\geq \frac{q^{k+1}-k q^k-q^{k-\delta+1}}{q^{k+1}}
		= 1 - \frac{k}{q} - q^{-\delta}
        \to 1
		\label{eq:asymptotic distance}
	\end{align}
	given the assumption on the scaling of $q$, $k$ and $\delta$ from \cref{eq:2RSscaling}.
	As we can see in \cref{eq:asymptotic distance}, we can satisfy the conditions to achieve identification capacity if we assume that $q$ grows to infinity faster that $k$.

	Identification based on single Reed-Solomon codes is also an option~\cite{MK06}, however they don't fulfil the  conditions to achieve capacity. A full size ($n=q$) single Reed-Solomon code would mean a $[n,\cdot,\cdot]_n$ block code which does not allow us to satisfy \cref{eq:optimal-tag}:
	\begin{align}
		&\lim_{n \to \infty} \frac{\log q(n)}{\log M(n)}  = \lim_{n \to \infty} \frac{\log n}{\log n} \to 1,
	\end{align}
	meaning that the tag is too big compared to the randomness. Shorter Reed-Solomon codes ($n<q$) can only make this limit larger.

	\subsection*{Example}
	We now proceed with an example of how such concatenated taging functions are built. For simplicity and clarity reasons, we will use the smallest possible code with the parameter set $(3,2,1)_{RS^2}=(3,2)_3^{RS} \circ (3^2,3)_{3^2}^{RS}$. The concatenated code is an
	\begin{equation}
		[q^{k+1}, kq^{k-\delta}, (q-k+1)(q^k-q^{k-\delta}+1)]_q = [27, 6, 14]_3
	\end{equation} 
	error-correction code.
	
	In order to obtain a tag from the concatenated Reed-Solomon code, we first need an identity. For example purposes, we pick the $587^{th}$ identity among the 729 total identities. In symbols of the field $F_{3^2}$, this is the string	
	\begin{equation}
		\underline{m} = 587 = 722_{F_{3^2}},
	\end{equation} 
	where the first string is the number in base $10$ while the second is the same number written in base $F_{3^2}$.
	Now, we can use the numbers 7, 2 and 2 as the orders of the field elements. We can order the elements of $F_{3^2} = \{0,\alpha^0, \alpha^1, \alpha^2, \alpha^3, \alpha^4, \alpha^5, \alpha^6, \alpha^7 \}$. With this order, we can rewrite our identity as
	\begin{equation}
		\underline{m} = 722_{F_{3^2}} = (\alpha^6 \ \alpha^1 \  \alpha^1) = (\alpha^6 \ \alpha \ \ \alpha)
	\end{equation}
	where $\alpha$ is the primitive element of $F_{3^2}$.
	Now we are ready to compute the tag function for the identity $\underline{m}$. 
	It is time to introduce the generator matrix of the outer code $(3^2,3)_{3^2}^{RS}$ with what we have seen in \cref{chap:preliminaries}:
	\begin{align}
		G_\mathrm{o} = \begin{pmatrix}
			\alpha^0 & \alpha^0 & \alpha^0 & \alpha^0 & \alpha^0 & \alpha^0 & \alpha^0 & \alpha^0 & \alpha^0\\
			0 & \alpha^0 & \alpha^1 & \alpha^2 & \alpha^3 & \alpha^4 & \alpha^5 & \alpha^6 & \alpha^7\\
			0 &\alpha^0 & \alpha^2 & \alpha^4 & \alpha^6 & \alpha^0 & \alpha^2 & \alpha^4 & \alpha^6
		\end{pmatrix}.
	\end{align}
	Multiplying $\underline{m}$ with $G_\mathrm{o}$, we get the outer codeword
	\begin{align}
		\underline{c}(\underline{m}) &= \underline{m} \cdot G_\mathrm{o} \\
		&= (\alpha^6 \ \alpha \ \ \alpha) \cdot G_\mathrm{o}\\
		&= (\alpha^6 \ \alpha^7  \ \alpha^3 \  0 \ 0 \  \alpha^6 \  \alpha^4 \  \alpha^3 \  \alpha^4) \in {F_{3^2}}^9.
	\end{align}
	
	Before calculating the final codeword, we need two preparation steps.
	First, we need to expand the symbols of the first codeword by transforming it's base field from $F_{3^2}$ to $F_3$. The reason is that the next Reed-Solomon code $(3^2,3)_3^{RS}$ is based on field $F_3$. The way to do this is through representing the symbols back in their string form. If we apply this to the first codeword in our example, the outer codeword will become:
	\begin{equation}
		\underline{c}(\underline{m}) = ((22) \ (21)  \ (12) \  (00) \ (00) \  (22) \  (20) \  (12) \ (20)) \in {F_{3}}^{18}.
	\end{equation}
	Now our codeword is ready to be concatenated with the inner code. But since a concatenation means the symbols will be encoded one by one, instead of encoding only the codeword, we need a larger generator matrix than the regular generator matrix of $(3,2)_3^{RS}$. We need the direct sum of this generator matrix with itself by $9$ times, as the second preparation step. The specified block sum looks like this:
	\begin{align*}
		G_B &
		= \bigoplus_{\ell=0}^8 G_\mathrm{i}
		= \bigoplus_{\ell=0}^8 
		\begin{pmatrix}
			1 & 1 & 1\\
			0 & 1 & 2
		\end{pmatrix}
	\end{align*}
	where $G_\mathrm{i}$	is the generator matrix of the \emph{inner} code $(3,2)_3^{RS}$. 
	Now the multiplication 
	\begin{equation}
		T_{\underline{m}} = \underline{c}(\underline{m}) \cdot G_B
	\end{equation}
	will yield the final codeword $T_{\underline{m}}$. Every symbol of the first codeword $c$ will be encoded with $G_\mathrm{i}$. Since each symbol now has two trits and $G_\mathrm{i}$ transforms two trits into three trits, we get
	\begin{equation}
		T_{587} = (2 \ 1 \ 0 \ 2 \ 0 \ 1 \ 1 \ 0 \ 2 \ 0 \ 0 \ 0 \ 0 \ 0 \ 0 \ 2 \ 1 \ 0 \ 2 \ 2 \ 2 \ 1 \ 0 \ 2 \ 2 \ 2 \ 2).
	\end{equation}
	At this point we have the tagging function of the $587^{th}$ identity. If, for instance, the corresponding tag of the randomness $j=5$ in the $i=587$ identity is required, we simply calculate this tagging function as done above and look at position $j=5$. 
	The preprocessing step is completed as we use the concatenated string  
	\[(j, T_i(j)) = (5, T_{587}(5)) = (5,1)\] 
	as encoding of the identity $i=587$, which must now be sent through a channel to the receiver.

	\section{Comparison of Identification and Transmission}
	
	Here we make a comparison between the identification and the transmission schemes at equal blocklengths. Both the single Reed-Solomon code and the double Reed-Solomon code constructions will be compared to their counterparts in the transmission scheme.
	Since we are coding for the noiseless channel, the number of bits sent is the size of the transmission scheme.
	We will thus be comparing the amount of bits used in the construction of the $(j, T_i(j))$ to the number of identities achieved by the error-correction code used for identification. 
	However, notice that while the identification code will have a certain false identification probability, the error in the transmission code is completely absent.
	Therefore part of the comparison is the trade-off between the introduced error and the increase in the number of identities.

	\subsection{Single Reed-Solomon Code}

	As already explained, the codewords of an error-correction code represent each identity. Thus there are as many identities as the number of possible codewords.
	In a $(q,k)_q^{RS}$ Reed-Solomon code, there are $q^k$ codewords and thus that many identities. Therefore the Single Reed-Solomon construction~\cite{MK06} has $q^k$ identification messages.
	The blocklength and the alphabet size of a $(q,k)_q^{RS}$ Reed-Solomon code are both $q$, therefore the transmission $(j,T_i(j))$ takes $q^2$ elements.
	If we use the $q^2$ elements to send $q^2$ messages via noiseless transmission, we would achieve a rate 
	\begin{align}
		r_{T}  = \frac{\log q^2}{n} = 2 \frac{\log q}{n},
	\end{align}
	where $n$ is now the blocklength of the noiseless channel. 
	In comparison the rate achieved by identification is 
	\begin{align}
		r_{ID}  = \frac{\log q^k}{n} = k \frac{\log q}{n}
	\end{align}
	
	Therefore we have an increase in the rate of 
	\begin{align}
		\frac{r_{ID}}{r_{T}}  = \frac{k}{2}
	\end{align}
	at the cost of introducing, in the worst case, a false identification error probability of 
	\begin{align}
		1 - \frac{d}{q} &= 1 - \frac{q-k+1}{q} = \frac{k-1}{q}.
	\end{align}
	
	As long as $q$ grows faster than $k$, in the asymptotic case where they both grow to infinity, we can increase the encoded identification messages and decrease maximum false identification probability $\lambda_2$. This shows that for the same blocklength, the single Reed-Solomon code construction has polynomially more messages than its transmission counterpart.
	
	By letting $k$ grow with the same order as $q$, an exponential increase in the rate is still possible~\cite{MK06}, however as we will see now, the double Reed-Solomon code achieves an increase in the rate that is exponential also in $k$.
	
	\subsection{Double Reed-Solomon Code}
	
	We now make the same comparison for the double Reed-Solomon code of \cref{eq:RS2}: we compare the number of identification messages with the size of the randomness/tag pairs.
	Recall that the double Reed-Solomon code is a 
	\begin{equation}
		(q,k,\delta)_{RS^2} \in [q^{k+1}, kq^{k-\delta}, (q-k+1)(q^k-q^{k-\delta}+1)]_q
	\end{equation} 
	There are thus $q^{kq^{k-\delta}}$ identification messages in this construction.
	
	The size of the preprocessed codeword $(j, T_i(j))$ is computed as follows. 
	This time there are $q^{k+1}$ possible randomness values with again $q$ possible tags, for a total of $q^{k+2}$. 
	The rate achieved by transmission is thus 
	\begin{align}
		r_{T} &
		= \frac{\log q^{k+2}}{n} 
		= (k+2) \frac{\log q}{n}.
	\end{align}
	In contrast, the rate achieved by identification is 
	\begin{align}
		r_{T} &
		= \frac{\log q^{kq^{k-\delta}}}{n} 
		= kq^{(k-\delta)} \frac{\log q}{n}.
	\end{align}
	The resulting increase in the rate is thus
	\begin{align}
		\frac{r_{ID}}{r_{T}}  
		= \frac{kq^{k-\delta}}{k+2} \approx q^{k-\delta}
		= \mathit{exp}\left( (k-\delta)n \frac{\log q}{n} \right)
		= \mathit{exp}\left( (k-\delta)n r_{T}  \right)
	\end{align}
	The rate of a double Reed-Solomon code construction is thus exponential to the rate achieved with simple transmission.
	
	The trade-off in the false identification error introduced to achieve this rate is given by the distance 
	\begin{align}
		d &= (q - k + 1)(q^k - q^{k-\delta} + 1)\geq (q - k)(q^k - q^{k-\delta} ) \geq q^{k+1} - kq^k- q^{k+1-\delta},
	\end{align}
	where we assumed that $k> 1$.
	This gives a maximum false identification probability of 
	\begin{align} \label{eq:false_alarm_concat}
		\lambda_2 &\leq 1 - \frac{d}{q^{k+1}}
		= 1 - \frac{q^{k+1} - kq^k- q^{k+1-\delta} }{q^{k+1}}\\
		&= \frac{k}{q} +  q^{-\delta} 
		\in O\left(\frac{k}{q}\right).
	\end{align}
	And thus we can still achieve the exponential increase in messages while still sending the error to zero.
	Notice, however, that the scaling at which the error goes to zero is unchanged and arguably slow compared to the scaling of the amount of identification messages.

	\subsection{Implementation and Simulation}
	\label{chap:simulation}
	The major part of our contribution lies in the implementation of the double Reed-Solomon code construction using Sagemath\cite{sagemath}.
	Coding the steps as presented so far is an inefficient solution. 
	In particular, producing the generator matrices is costly in memory and a waste of computation.
	Already at small parameters $(7,5,2)_{RS^2}$ the generator matrices become too large to handle. 
	Even storing only one codeword is prohibitive.
	The only advantage to such a solution would be the instant access to the precomputed tags which is nullified by the impossibility of taking advantage of the doubly exponential growth of the codewords.
	
	Instead, the only way of taking advantage of the doubly exponential growth is to compute each tag on demand, using the Reed-Solomon codes as polynomial evaluations as per the initial definition (see \cref{subsect:rs}), and computing single symbols of each codeword at every transmission of an encoded  identity. 
	In the final implementation, a tag is computed as follows:	
	\begin{itemize}
		\item Divide the randomness $j$, ranging from $0$ to $q^{k+1} - 1$ by $q$. The quotient $j \div q$ shows us which column of the generator matrix of the $(q^k,q^{k-\delta})_{q^k}^{RS}$ we need to use. We don't need the other columns to calculate the necessary tag. Using only that column we will get one symbol $\tilde t$ in the alphabet of size $q^k$. The remainder of the division, $j \mod q$ is used later.
		\item Expand the symbol $\tilde t$ into $k$ symbols of size $q$. The list of these $q$ elements will be called the expanded codeword.
		\item Find the column - or the code locator - in the generator matrix of $(q,k)_q^{RS}$ with the index as the remainder $j \mod q$ and multiply the expanded codeword with that column scalarly, or simply evaluate it with the picked code locator.
	\end{itemize}
	
	The result of this last evaluation gives us the necessary tag. This method saves us a lot of memory compared to using the generator matrices. 
	The bottleneck of the implementation at this point is listing the elements of the extension field $F_{q^k}$, which renders parameters of the order $(11,8,4)_{RS^2}$ again intractable.
	Ideally we would want to index the desired field element as a parameter of the field and obtain the desired element. This technique works on prime fields $F_q$, but not on extension fields $F_{q^k}$. On prime fields $F_q(129)$ returns $129 \mod q$, which is correct, however on extension fields $F_{q^k}(129)$ also returns $129 \mod q$, which is not correct for $q \leq 129$.
	
	Further improvements were obtained by changing the way field elements are generated.
	The "next()" method of the field class does what its name suggests: it gives the next field element. 
	In this way, we can generate the elements by sequentially producing the next element from the zero element. If, for example, we wanted the $10234^{th}$ element of the field, we would start from the $0^{th}$ element of the field and using the "next()" method 10234 times to get to the $10234^{th}$ element of the field. 
	With this method, the computation time of an element remains feasible up to elements of the order $10^8$, with elements larger that $10^9$ taking hours of computation time.
	
	A final improvement was made again, changing the generations method.
	In \cref{chap:preliminaries}, we see that the primitive element of a field can create the field from scratch by its exponents, where $0$ and $\alpha^0$ would be the first two elements and $\alpha^{q^k - 2}$ would be the last element, if we are speaking of the field $F_{q^k}$. 
	Generating the elements with this method resulted in several order of magnitude of improvement in the computation time.
	With this final method, the boundary of the feasible computations lay at parameters $(17,12,6)_{RS^2}$ for which the computation of a single tag took $\sim1.5$ hours.

	At this point the achieved number of identities are far beyond the ones achievable with transmission, however the maximum false identification ratio $\lambda_2 \approx k/q = {12}/{17}\approx 60\%$ is too high to be acceptable. 
	In order, to bring the error down, the base field size $q$ must grow much larger than the parameter $k$. 
	In the final simulations we decided to keep fixed $k=3$ and $\delta=2$ and simply increase the field size $q$, studying the performance of only $(q,3,2)_{RS^2}$ codes. 
	Recall that the increase in rate is 
	\begin{align}
		\frac{r_{ID}}{r_{T}}  
		= \mathit{exp}\left( (k-\delta)n r_{T}  \right).
	\end{align}
	By fixing the parameter $k$ we did not achieve identification capacity but we still hit an exponential increase in rate compared to transmission.
	This also allowed for a vast increase in the range of computable values for $q$, and reduced the false identification error as rapidly as we could increase $q$.
	The limiting value in this regime is $q\approx 10^8$ which provides a false identification error of $\lambda_2 = 10^{-8}$, at the cost of $\approx 2.5$ hours required to calculate one tag.
	In the next section we present these results in detail.

	\section{Results}
	\label{chap:results}
	
	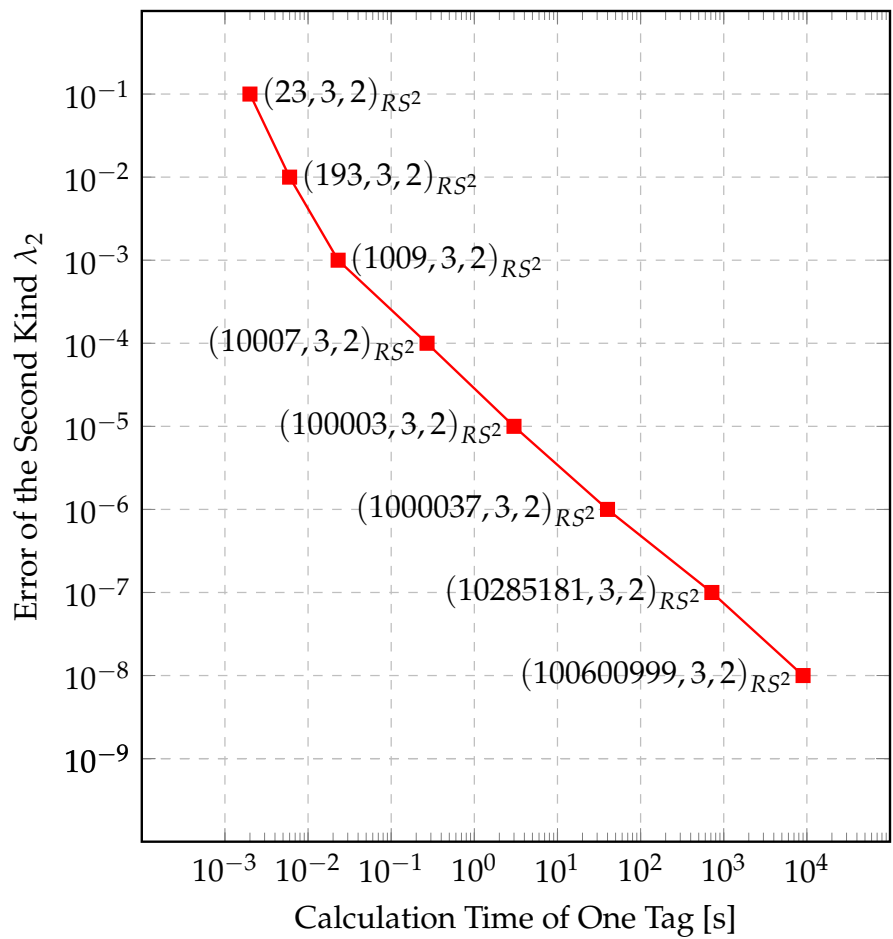
\begin{figure} 
		\centering
		
		\begin{tikzpicture}[scale=1.2]
		\begin{axis}[
		xmode=log,
		ymode=log,
		title={},
		xlabel={Calculation Time of One Tag [s]},
		ylabel={Error of the Second Kind $\lambda_2$},
		xmin=0.0001, xmax=100000,
		ymin=0.0000000001, ymax=1,
		x=0.4cm,
		y=0.4cm,
		xtick={0.001,0.01,0.1,1,10,100, 1000, 10000},
		ytick={0.000000001, 0.000000001, 0.00000001, 0.0000001, 0.000001, 0.00001, 0.0001, 0.001, 0.01, 0.1},
		legend pos=north west,
		ymajorgrids=true,
		xmajorgrids=true,
		yminorticks=false,
		yminorticks=true,
		grid style=dashed,
		tick align=inside,
		]
		
		\addplot[
		color=red,
		mark=cube*,
		]
		coordinates {
			(9000, 0.00000001)(720, 0.0000001)(40, 0.000001)(3, 0.00001)(0.27, 0.0001)(0.023, 0.001)(0.006, 0.01)(0.002, 0.1)
		};
		\node at (axis cs:0.002, 0.1) [anchor=west] {$(23,3,2)_{RS^2}$};
		\node at (axis cs:0.006, 0.01) [anchor=west] {$(193,3,2)_{RS^2}$};
		\node at (axis cs:0.023, 0.001) [anchor=west] {$(1009,3,2)_{RS^2}$};
		\node at (axis cs:0.27, 0.0001) [anchor=east] {$(10007,3,2)_{RS^2}$};
		\node at (axis cs:3, 0.00001) [anchor=east] {$(100003,3,2)_{RS^2}$};
		\node at (axis cs:40, 0.000001) [anchor=east] {$(1000037,3,2)_{RS^2}$};
		\node at (axis cs:720, 0.0000001) [anchor=east] {$(10285181,3,2)_{RS^2}$};
		\node at (axis cs:9000, 0.00000001) [anchor=east] {$(100600999,3,2)_{RS^2}$};	
		\end{axis}
		\end{tikzpicture}
		\caption{Trade-off between complexity and reliability.}
		\label{fig:trade-off}
	\end{figure}
	
	We used $=3k$ and $\delta=2$ at all times in order for $q$ to grow larger than $k$.
	To reach channel capacity $k$ would also need to grow, however, making both $q$ and $k$ grow together makes the simulation prohibitive, requiring a much greater RAM in the simulation computer.
	
	Regardless of the algorithm used, the time requirement decreases with the complexity, in turn increasing the error of the second kind $\lambda_2$. 
	If a false identification error lower than $10^{-5}$ is needed, then with our algorithm this implies a calculation time of at least $1\mathrm{s}$. And vice versa; requiring the calculation time to be under $1\mathrm{ms}$, we must allow a smaller number of identities and a false identification error of at least 10\%. 
	These results are displayed in \cref{fig:trade-off}.
	Simply said, complexity reduces the maximum false identification probability at the cost of calculation time. Note that this calculation time involves everything from building the system with the selected parameters, picking a random identity among all identities and picking the randomness to calculate the corresponding tag.

	\begin{figure} 
		\centering
		
		\begin{tikzpicture}[scale=1.2]
		\begin{axis}[
		xmode=log,
		ymode=log,
		title={},
		xlabel={Calculation Time of One Tag [s]},
		ylabel={Logarithm of Number of Identities},
		xmin=0.0001, xmax=100000,
		ymin=10, ymax=10e10,
		x=0.4cm,
		y=0.4cm,
		xtick={0.001,0.01,0.1,1,10,100, 1000, 10000},
		ytick={10, 100,1000,10000,100000,1000000,10000000, 100000000, 1000000000, 10000000000},
		legend pos=north west,
		ymajorgrids=true,
		xmajorgrids=true,
		grid style=dashed,
		tick align=inside,
		]
		
		\addplot[
		color=violet,
		mark=cube*,
		]
		coordinates {
			(9000, 2400000000)(720, 210000000)(40, 18000000)(3, 1500000)(0.27, 120000)(0.023, 9000)(0.006, 1380)(0.002, 93)
		};
		\node at (axis cs:0.002, 93) [anchor=west] {$(23,3,2)_{RS^2}$};
		\node at (axis cs:0.006, 1380) [anchor=west] {$(193,3,2)_{RS^2}$};
		\node at (axis cs:0.023, 9000) [anchor=west] {$(1009,3,2)_{RS^2}$};
		\node at (axis cs:0.27, 120000) [anchor=east] {$(10007,3,2)_{RS^2}$};
		\node at (axis cs:3, 1500000) [anchor=east] {$(100003,3,2)_{RS^2}$};
		\node at (axis cs:40, 18000000) [anchor=east] {$(1000037,3,2)_{RS^2}$};
		\node at (axis cs:720, 210000000) [anchor=east] {$(10285181,3,2)_{RS^2}$};
		\node at (axis cs:9000, 2400000000) [anchor=east] {$(100600999,3,2)_{RS^2}$};	
		\end{axis}
		\end{tikzpicture}
		\caption{The relationship between the number of identities and the calculation time of one tag.}
		\label{fig:iden-numb-calc-time}
	\end{figure}
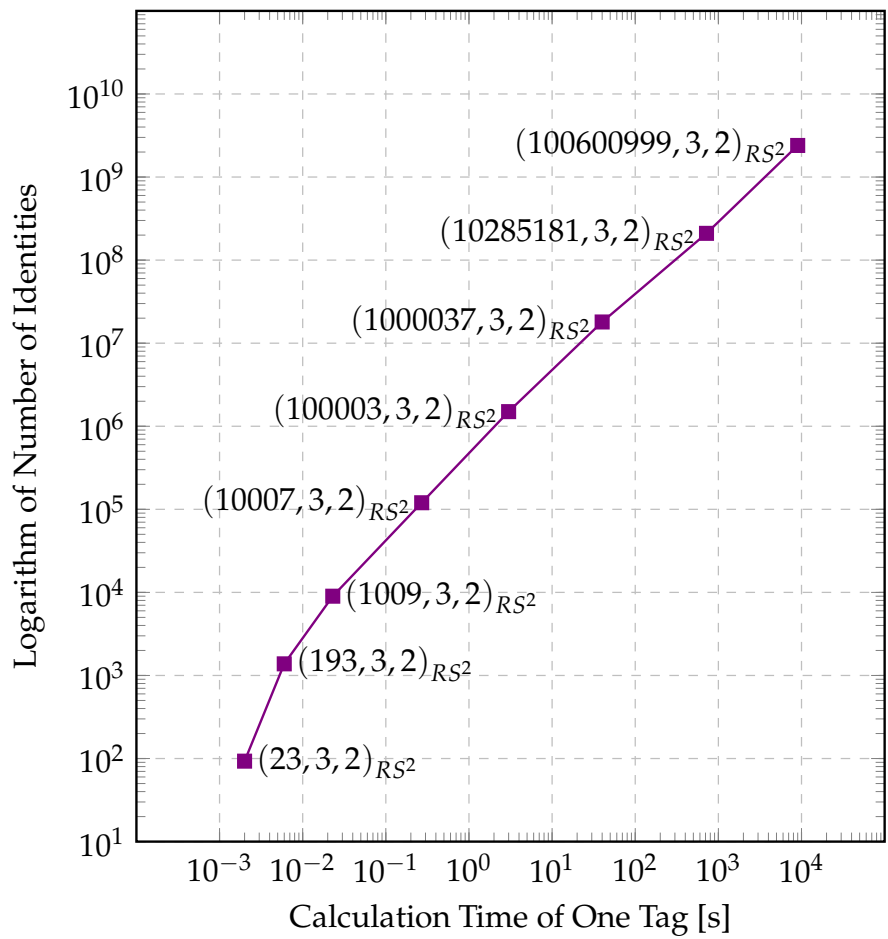
	
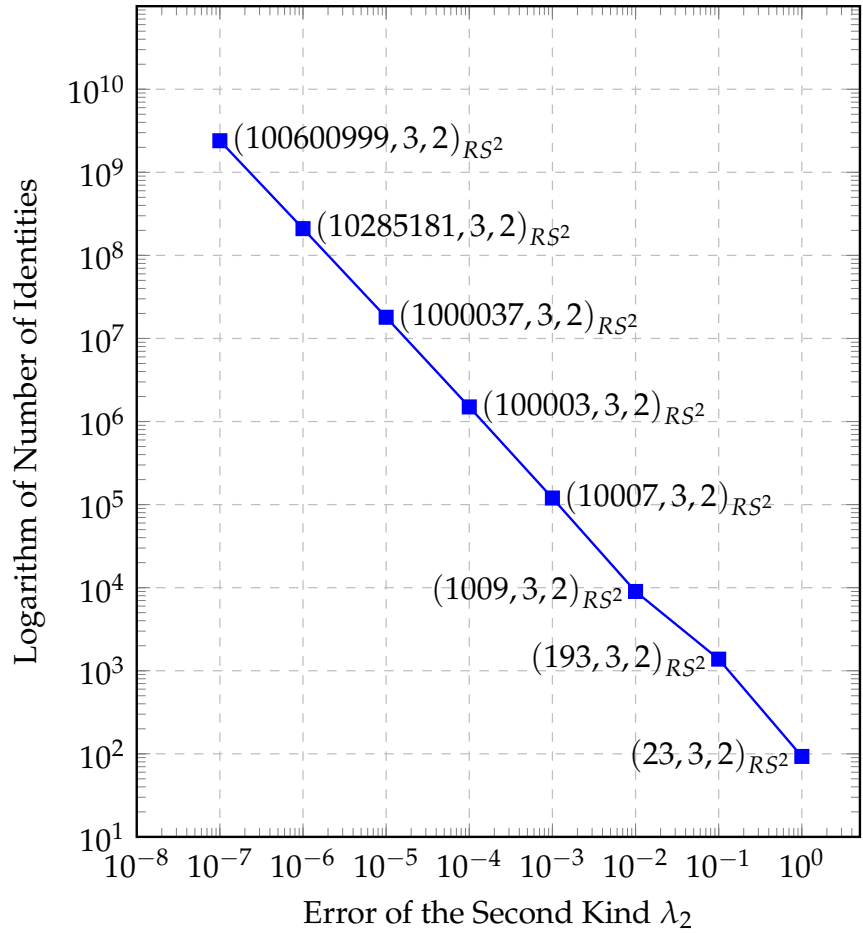
\begin{figure} 
		\centering
		
		\begin{tikzpicture}[scale=1.2]
		\begin{axis}[
		xmode=log,
		ymode=log,
		title={},
		xlabel={Error of the Second Kind $\lambda_{2}$},
		ylabel={Logarithm of Number of Identities},
		xmin=10e-9, xmax=5,
		ymin=10, ymax=10e10,
		x=0.4cm,
		y=0.4cm,
		xtick={10e-9,10e-8,10e-7,10e-6,10e-5,10e-4, 10e-3, 10e-2, 10e-1},
		ytick={10, 100,1000,10000,100000,1000000,10000000, 100000000, 1000000000, 10000000000},
		legend pos=north west,
		ymajorgrids=true,
		xmajorgrids=true,
		grid style=dashed,
		tick align=inside,
		]
		
		\addplot[
		color=blue,
		mark=cube*,
		]
		coordinates {
			(10e-8, 2400000000)(10e-7, 210000000)(10e-6, 18000000)(10e-5, 1500000)(10e-4, 120000)(10e-3, 9000)(10e-2, 1380)(10e-1, 93)
		};
		\node at (axis cs:10e-8, 2400000000) [anchor=west] {$(100600999,3,2)_{RS^2}$};
		\node at (axis cs:10e-7, 210000000) [anchor=west] {$(10285181,3,2)_{RS^2}$};
		\node at (axis cs:10e-6, 18000000) [anchor=west] {$(1000037,3,2)_{RS^2}$};
		\node at (axis cs:10e-5, 1500000) [anchor=west] {$(100003,3,2)_{RS^2}$};
		\node at (axis cs:10e-4, 120000) [anchor=west] {$(10007,3,2)_{RS^2}$};
		\node at (axis cs:10e-3, 9000) [anchor=east] {$(1009,3,2)_{RS^2}$};
		\node at (axis cs:10e-2, 1380) [anchor=east] {$(193,3,2)_{RS^2}$};
		\node at (axis cs:10e-1, 93) [anchor=east] {$(23,3,2)_{RS^2}$};	
		\end{axis}
		\end{tikzpicture}
		\caption{The relationship between parameter sets and $\lambda_2$.}
		\label{fig:lambda2-parametersets}
	\end{figure}

	A similar trade-off exists with the number of identification messages.
	We saw that there is an exponential relationship between the calculation time of one tag and the total number of identities. When we have a tenfold calculation time of one tag, the exponent of the number of identities becomes tenfold. For examples, the reader can see \cref{fig:iden-numb-calc-time}. These results\footnote{Note that in both\cref{fig:iden-numb-calc-time,fig:lambda2-parametersets}, the vertical axis expresses the common logarithm of number of identities, but not the number of identities themselves. This was the only way we could fit a double exponential growth in a plot.} show us that we can increase the total number of identities exponentially if we can endure a linear increase of the calculation time, depending on the applications.
	
	Another simulation yielded us the relationship between the total number of identities and the error of the second kind. The thing to notice here, is that when the $q$ grows ten-fold, the error of the second kind becomes 10\% of the previous value. This is no surprise, since we fixed $k$ to 3 and $\delta$ to 2:
	
	\begin{align}
		\lambda_2 &= 1 - \frac{d}{n}
		= 1 - \frac{(q-k+1)(q^k - q^{k-\delta}+1)}{q^{k+1}}
		\\&
		= 1 - \frac{(q-2)(q^3-q+1)}{q^4}
		= \frac{2}{q} + \frac{1}{q^2} - \frac{3}{q^3} + \frac{2}{q^4}
		\\&
		\in O\left( \frac{1}{q}\right) \label{eq:1/q}	
	\end{align}
	
	As we see in \cref{eq:1/q}, for large $q$, the error of the second kind equals, in our example set with $(q,3,2)_{RS^2}$, approximately $\frac{1}{q}$, which is why the error of the second kind $\lambda_2$ and $q$ seem inversely proportional. They are indeed inversely proportional. This brings us to the conclusion that lowering the error probability is expensive and the user of the system needs to prioritize between a slow but mostly accurate system, a fast system which has non-neglectable errors, or a balanced system. Our simulation results can be seen in \cref{fig:lambda2-parametersets}.
	
	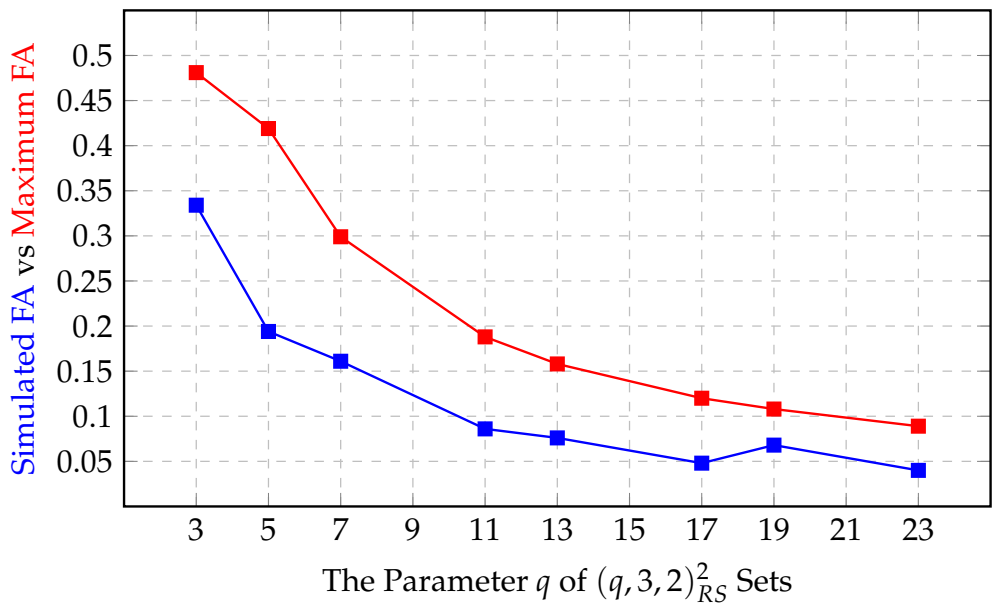
\begin{figure} 
		\centering
		
		\begin{tikzpicture}[scale=1.2]
		\begin{axis}[
		scaled ticks=false, 
		tick label style={/pgf/number format/fixed},
		xmode=linear,
		ymode=linear,
		title={},
		xlabel={The Parameter $q$ of $(q,3,2)_{RS}^2$ Sets},
		ylabel={\textcolor{blue}{Simulated FA} vs \textcolor{red}{Maximum FA}},
		xmin=1, xmax=25,
		ymin=0, ymax=0.55,
		x=0.4cm,
		y=10cm,
		xtick={3,5,7,9,11,13, 15, 17, 19, 21, 23},
		ytick={0.05,0.1, 0.15,0.2,0.25,0.3,0.35,0.4,0.45,0.5},
		legend pos=north west,
		ymajorgrids=true,
		xmajorgrids=true,
		grid style=dashed,
		tick align=inside,
		]
		
		\addplot[
		color=blue,
		mark=cube*,
		text="abdurrahman"
		]
		coordinates {
			(3, 0.334)(5, 0.194)(7, 0.161)(11, 0.086)(13, 0.076)(17, 0.048)(19, 0.068)(23, 0.04)
		};
		
		\addplot[
		color=red,
		mark=cube*,
		]
		coordinates {
			(3, 0.481)(5, 0.419)(7, 0.299)(11, 0.188)(13, 0.158)(17, 0.12)(19, 0.108)(23, 0.089)
		};
		\end{axis}
		\end{tikzpicture}
		\caption{Another simulation by fixing an identity and the randomness.}
		\label{fig:fixedrandomness}
	\end{figure}
	
	We also ran a different simulation: we fix a randomly chosen identity and a randomness value, then we start to pick different identities at random and compare their tags in the same position with the tag of our original identity at the fixed position. We repeat this comparison with 1000 different secondary identities and set the said "false identification ratio" as the number of recognized false identifications over 1000 iterations.
	This simulation can be seen in \cref{fig:fixedrandomness}. It is assuring that no false identification ratio in the simulation exceeds the theoretical limits\footnote{Note that in \cref{fig:fixedrandomness}, the 3 in the horizontal axis corresponds to a parameter set of $(3,2,1)_{RS^2}$ exceptionally.}.
	
	
	One could make two remarks here. First, that the simulated false identification ratio values are approximately inversely proportional to $q$. This is no surprise, since the originally-chosen tag is a fixed number from the set $\{0,1,2,...,q-1\}$. And yet, the secondary identity's tag at the same position will also be another number from the same set $\{0,1,2,...,q-1\}$. The broad probability that a position possesses a certain value is therefore $\frac{1}{q}$, assuming that each possible tag comes up almost the same number of times, which also seems like the case in this last simulation of ours. So it is no coincidence that we meet the same number in the same position of other identities with the probability $\frac{1}{q}$.
	
	The second remark, which is the most important aspect here, is that the simulated false identification ratio is an average error probability, unlike this paper focusing on the maximum error probability. 
	
	\section{Conclusions}
	
	This paper shows that identification can efficiently reach many more messages than transmission at the cost of a manageable additional error and the inability to decode all messages. We can see from the results, that making the maximum probability of the error of the second kind $\lambda_2$ smaller also means a trade-off in the computation time, which contributes to the latency of the scheme. So if one wants smaller errors, our paper shows that one needs a more powerful system to calculate the tags under an acceptable latency.
	We believe that some improvement in the computational performance is still possible, which is left for future work, together with an analytic computation of the computational complexity of computing the tags.
	
	
	For future works on identification using tag codes, adding some security element into play is important to protect the information from being eavesdropped. The benefit of protecting a tag code is that we only need to protect the tag part of the codeword, because the randomness is uncorrelated with the identity, so it carries no useful information and  will not increase the success probability of the eavesdropper.
	
	Furthermore, most of the channel capacity in an \textit{identification} code is used to produce common randomness on which to compute the tag. 
	In systems where the common randomness is given as a resource, or can be synchronized and produced on demand by other means, the resulting uses of the channel diminish drastically. 
	Then, rather than increasing the size of the messages, an advantage can be achieved by keeping the number of messages comparable and instead exponentially or even doubly exponentially reducing the number of channel uses. 
	In such a scenario we can think of identification as converting the latency of communication into latency of computation, which might become advantageous in applications where the increased number of messages is not the main interest.

	\funding{Christian Deppe and Roberto Ferrara  were supported by the Bundesministerium f\"ur Bildung und Forschung (BMBF) through the grant 16KIS1005.}

	\reftitle{References}
	

\end{document}